\documentclass[12pt]{article}

\usepackage{graphicx}
\usepackage{dcolumn}
\usepackage{bm}
\usepackage{amsmath}
\usepackage{epstopdf}
\usepackage{amsfonts}
\usepackage{amssymb}
\usepackage{epsfig}
\usepackage{tabularx}
\usepackage{color}

\usepackage[colorlinks=true,citecolor=blue,urlcolor=magenta,breaklinks]{hyperref}

\newcommand{\be}{\begin{equation}}
\newcommand{\en}{\end{equation}}
\newcommand{\bea}{\begin{eqnarray}}
\newcommand{\ena}{\end{eqnarray}}

\begin{document}

\begin{titlepage}

\centerline{\large \bf {Scale-dependent slowly rotating black holes with flat horizon structure}}

\vskip 1cm

\centerline{\'Angel Rinc{\'o}n}

\vskip 1cm

\centerline{Instituto de F{\'i}sica, Pontificia Universidad Cat{\'o}lica de Valpara{\'i}so,} \centerline{Avenida Brasil 2950, Casilla 4059, Valpara{\'i}so, Chile.}

\vskip 1cm

\centerline{Grigoris Panotopoulos}

\vskip 1cm

\centerline{Centro de Astrof{\'i}sica e Gravita\c c\~ao-CENTRA, Departamento de F{\'i}sica,} 

\centerline{Instituto Superior T{\'e}cnico-IST, Universidade de Lisboa-UL,}

\centerline{Avenida Rovisco Pais 1, 1049-001, Lisboa, Portugal.}

\vskip 1.5cm

\centerline{email:
\href{mailto:angel.rincon@pucv.cl}
{\nolinkurl{angel.rincon@pucv.cl}}
}

\vskip 0.5cm

\centerline{email:
\href{mailto:grigorios.panotopoulos@tecnico.ulisboa.pt}
{\nolinkurl{grigorios.panotopoulos@tecnico.ulisboa.pt}}
}

\vskip 1.5cm

\begin{abstract}
We study slowly rotating four-dimensional black holes with flat horizon structure in scale-dependent gravity. First we obtain the solution, and then we study thermodynamic properties as well as the invariants of the theory. The impact of the scale-dependent parameter is investigated in detail. We find that the scale-dependent solution exhibits a single singularity at the origin, also present in the classical solution.
\end{abstract}

\end{titlepage}

%%%%%%%%%%%%%%%%%%%%%%%
\section{Introduction}
%%%%%%%%%%%%%%%%%%%%%%

Einstein's General Relativity (GR) \cite{GR} not only is very successful, as by now it has passed numerous tests \cite{tests1,tests2,tests3}, but at the same time is considered to be one of the most beautiful theories ever formulated. As successful as it may be as a classical theory, at quantum level technically speaking GR falls into the class of non-renormalizable theories. Although by now we know how to extract quantum predictions from a non-renormalizable theory using the techniques of effective field theory (to which GR fits perfectly) \cite{Donoghue:1994dn}, the problem remains. 
The formulation of a consistent theory of quantum gravity remains an open issue in modern theoretical physics. Currently in the literature there are several approaches that have been proposed and studied (see for instance\cite{QG1,QG2,QG3,QG4,QG5,QG6,QG7,QG8,QG9} and references therein). A closer look reveals that all of those share the same property, namely the fact that the basic quantities that enter into the defining action of our favorite model, such as the cosmological constant, the gravitational or electromagnetic coupling etc, become scale dependent (SD) quantities. 
This was to be expected, as it is well-known that a generic feature of standard quantum field theory is the scale dependence at the level of the effective action.

\smallskip

Black holes (BHs),  a remarkable prediction of all metric theories of gravity, are exciting and important objects for theories of gravity, both at the classical and quantum level, linking together several scienticic areas and research fields, from astrophysics and gravitation to statistical mechanics and quantum physics. There are currently three main classes of BHs, namely astrophysical, primordial and mini-BHs. In particular, during the final stages of gravitational collapse of massive stars astrophysical BHs emerge, density inhomogeneities in the early Universe seed the formation of primordial BHs, while mini-BHs are expected to form at colliders or in the atmosphere of the earth in TeV-scale gravity scenarios in D-brane constructions of the Standard Model \cite{miniBH1,miniBH2,miniBH3}. The first image of a black hole shadow announced last year by the Event Horizon Telescope \cite{L1,L2,L3,L4,L5,L6}, and also the numerous direct detection by the LIGO/Virgo collaborations \cite{ligo1,ligo2,ligo3,ligo4,ligo5} of gravitational waves from BH binary systems, have established the existence of BHs over the last 5 years or so.

\smallskip

Within the framework of GR the most general BH solution is the Kerr-Newman geometry characterized by its mass, angular momentum and electric charge, see e.g. \cite{solutions}. Since, however, astrophysical BHs are expected to be electrically neutral, the most interesting cases to be considered are either the Schwarzschild \cite{SBH} or the Kerr geometry \cite{kerr}. Rotating BH solutions have attracted a lot of attention recently due to the great interest of the community in BH shadows, see e.g. \cite{synge,luminet,Bambi:2008jg,Bambi:2010hf,study1,study2,Moffat,carlos1,quint2,carlos2,
study3,study4,study5,bobir2017,study6,sudipta2019,shakih2019b,Konoplya:2019sns,chicos} and references therein. Furthermore, the current cosmic acceleration \cite{SN1,SN2} as well as the AdS/CFT correspondence \cite{adscft1,adscft2} motivate the study of space-times with a non-vanishing cosmological constant (CC). The Ba{\~n}ados-Teitelboim-Zanelli black hole \cite{BTZ1,BTZ2}, which marked the interest in lower-dimensional gravity, is sourced by a negative cosmological constant, while BH solutions with flat horizon structure (cylindrical, planar or toroidal) \cite{lemos1,lemos2,lemos3}, also require a negative cosmological constant. 

\smallskip

So far the impact of the scale-dependent gravity on Cosmology, relativistic stars as well as BH physics has been studied over the last years \cite{SD0,SD1,SD2,SD3,SD4,SD5,SD6,SD7,SD8,SD9,SD10}. To the best of our knowledge, however, the impact of the SD scenario on four-dimensional rotating BH solutions has not been studied yet. Therefore, in the present work we propose to obtain for the first time rotating BHs with flat horizon structure in the scale-dependence scenario.

\smallskip

The plan of this work is the following: In the next section we briefly describe the formalism. In section 3 we discuss four-dimensional rotating BHs with flat horizon structure in scale-dependent gravity, while their properties are investigated in the fourth section, where we discuss the thermodynamics as well as the invariants of the theory. Finally, we finish with some concluding remarks in the section 5. We work in geometrical units where $c=1=G_0$, and we adopt the mostly positive metric signature (-,+,+,+).

%%%%%%%%%%%%%%%%%%%%%%%%%%%%%%%%%%%%%%%%%%%%%%%%%%%%%%%
\section{Scale-dependent gravity in black hole physics}
%%%%%%%%%%%%%%%%%%%%%%%%%%%%%%%%%%%%%%%%%%%%%%%%%%%%%%%

In this section, we briefly review the main idea and the formalism of the SD gravity following \cite{SD9}. The motivation of the approach where the coupling constants evolve with a certain arbitrary scale is only understood in quantum gravity \cite{Rovelli:2007uwt}. Up to now 
a ``consistent and predictive" description of quantum gravity is still an open task in theoretical physics. Although one of the most popular approaches to quantum gravity is the so-called Loop Quantum Gravity (LQG), Exact Renormalization Groups (ERG) has recently attracted more adepts. The latter is precisely the inspiration of the scale-dependent approach, as the ERG technique starts from an average effective action with running couplings to incorporate quantum corrections. Thus, quantum effects are taken into account via the running of the coupling constants. It is essential to point out that the ERG and the SD scenario allow us to derive the equations for running couplings of an average effective action exactly, i.e., without the need for expansion of the couplings in powers of some small parameter. 

\smallskip

The scale-dependent scenario allows us to extend classical BH solutions to include quantum features that are assumed to be small. In the simplest case (without the presence of matter), we only have two couplings: i) Newton's constant $G_k$ and ii) the cosmological constant $\Lambda_k$. As usual, we can define an auxiliary parameter $\kappa_k \equiv 8 \pi G_{k}$. What is more, we have two extra fields, i.e., the metric tensor $g_{\mu \nu}$ and the arbitrary renormalization scale $k$.

\smallskip

The effective action $\Gamma[g_{\mu \nu}, k]$ is then written as \cite{SD9}
\begin{equation}
	\Gamma[g_{\mu \nu}, k] \equiv \int \mathrm{d}^4 x \sqrt{-g}
	\Bigg[ 
	\frac{1}{2 \kappa_k} \Bigl(\mathcal{R} - 2 \Lambda_k \Bigl) \ + \ \mathcal{L}_M
	\Bigg],
\end{equation}
where $\mathcal{L}_M$ is the Lagrangian density of the matter fields (if any), $g$ is the determinant of the metric tensor $g_{\mu \nu}$, $\mathcal{R}$ is the corresponding Ricci scalar, $\Lambda_k$ is the scale-dependent cosmological constant (CC), and $\kappa_k$ is the scale-dependent gravitational coupling. The average effective action variation with respect to the metric tensor gives rise to the effective Einstein's field equations:
\begin{equation}
G_{\mu \nu } + \Lambda_k g_{\mu \nu} \equiv \kappa_k T_{\mu \nu}^{\text{effec}},
\end{equation}
where the effective energy-momentum tensor is defined by
\begin{equation}
\kappa_k T_{\mu \nu}^{\text{effec}} =  \kappa_k T_{\mu \nu}^{M} - \Delta t_{\mu \nu}.
\end{equation}
It is mandatory to point out that the effective energy-momentum tensor takes into account two contributions: i) the usual matter content and ii) the non-matter source (provided by the running of the gravitational coupling), which is given by \cite{SD9}:
\begin{equation}
\Delta t_{\mu \nu} \equiv G_k \Bigl( g_{\mu \nu} \square - \nabla_{\mu} \nabla_{\nu} \Bigl) G_k^{-1}. 
\end{equation}
As already mentioned, the goal of this article is to investigate the properties of four-dimensional scale-dependent black holes with flat horizon structure, sourced by a negative cosmological constant only. Therefore, we set $T_{\mu \nu}^{M} = 0$ in the following, although, in principle, matter fields, such as electromagnetic sources, are always an exciting and vital ingredient in gravitational theories.

\smallskip

Next, varying the average effective action with respect to the additional field $k(x)$, we obtain an auxiliary equation to close the system of equations. The last condition reads:
\begin{equation}\label{eomk}
\frac{\delta \Gamma[g_{\mu \nu},k]}{\delta k} = 0.
\end{equation}
This restriction is usually considered as a posteriori condition towards background independence \cite{Stevenson:1981vj,Reuter:2003ca,Becker:2014qya,Dietz:2015owa,Labus:2016lkh,Morris:2016spn,Ohta:2017dsq}.
Taking advantage of (\ref{eomk}) we find a direct connection between $G_k$ and $\Lambda_k$ (or other couplings). Thus, we also notice that the cosmological constant is required to obtain self-consistent scale-dependent solutions. Otherwise, we should add additional matter Lagrangians into the action to maintain a consistent solution. So, the system is indeed closed after including the above equation.

%%%%%%%%%%%%%%%%%%%%%%%%%%%%%%%%%%%%%%%%%%%%%%%%%%%%%%%%%%%%%%%%%%%
\section{Scale-dependent black holes with flat horizon structure}
%%%%%%%%%%%%%%%%%%%%%%%%%%%%%%%%%%%%%%%%%%%%%%%%%%%%%%%%%%%%%%%%%%

In this section we obtain the scale-dependent solution, while its properties are
discussed in the next section.

%%%%%%%%%%%%%%%%%%%%%%%%%%%%%%%%%%%%%%
\subsection{Classical BH solutions}
%%%%%%%%%%%%%%%%%%%%%%%%%%%%%%%%%%%%%%

Let us consider the classical solution first. The starting point is Einstein's field equations without the presence of matter fields, and with a non-vanishing CC, $\Lambda=\pm 3/L^2$,  where $L > 0$ is a parameter with dimensions of length
\begin{equation}
G_{\mu \nu } + \Lambda g_{\mu \nu} = 0.
\end{equation}
The line element for the metric tensor without rotation has the general form 
\begin{equation}
ds^2 = -f_0(r) dt^2 + f_0(r)^{-1} dr^2 + r^2 \gamma_{ij} dx^i dx^j
\end{equation}
where $\gamma_{ij} dx^i dx^j$ represents the line element of a two-dimensional surface with constant curvature $k=-1,0,1$, and the indices $(i,j)=1,2$. The well-known solutions of General Relativity, such as the Schwarzschild and the Reissner-Nordstr{\"o}m geometries, correspond to spherical horizon structure where $k=1$. In this work, however, we shall consider solutions with a flat horizon structure where $k=0$.

\smallskip

For cylindrical or toroidal solutions we adopt a coordinate system $t,r,\phi,z$,
and for non-rotating solutions we make for the line element the ansatz \cite{lemos3}
\begin{equation}
ds^2 = -f_0(r) dt^2 + f_0(r)^{-1} dr^2 + r^2 \left( d \phi^2 + \frac{dz^2}{L^2} \right),
\end{equation}
where $0 < \phi < 2 \pi$, while the range of the $z$ coordinate determines the kind of the black hole, namely \cite{lemos3}.
\begin{equation}
- \infty < z < \infty, \, \, \, \, \, \, \, \, \textrm{cylindrical BH}
\end{equation}
\begin{equation}
0 < \frac{z}{L} < 2 \pi, \, \, \, \, \, \, \, \, \textrm{toroidal BH}.
\end{equation}
Given the field equations and the ansatz for the metric tensor, it is straightforward to obtain the expression for the lapse function, which is found to be \cite{lemos3}
\begin{equation}
f_0(r) = -\frac{4 M L}{r} - \frac{\Lambda r^2}{3} ,
\end{equation}
where $M > 0$ is the mass of the BH. Clearly, the existence of an event horizon requires a negative CC. Therefore, from now on we set $\Lambda = -3/L^2$, and the lapse function takes the form
\begin{equation}
f_0(r) = -\frac{\mu}{r} + \frac{r^2}{L^2},
\end{equation}
where we set $\mu=4 M L$, with $\mu > 0$ being a mass parameter proportional to the mass of the BH.

\smallskip

The lapse function may be rewritten in terms of the classical event horizon, $r_0$, as follows:
\begin{align}
f_0(r) &=  \bigg(\frac{r}{L}\bigg)^2 \Bigg[1 - \bigg(\frac{r_0}{r}\bigg)^3 \Bigg],
\end{align}
where $r_0$ is computed solving the algebraic equation $f_0(r_0) = 0$, and it is found to be
\begin{align}
r_0^3 &= \mu L^2 .
\end{align}

\smallskip

Next, the full rotating solution with angular velocity $\omega$ has been obtained in \cite{lemos2}, and in the slowly rotating limit it is given by
\begin{equation}
ds^2 = -f_0(r) dt^2 + f_0(r)^{-1} dr^2 + r^2 d \phi^2 + \frac{r^2}{L^2} dz^2 - 2 \: \left( \frac{\omega \mu L^2}{r} \right) \: dt d \phi ,
\end{equation}
where now there is a non-diagonal term proportional to $\omega$.

\smallskip

The thermodynamics is discussed in detail later on, see section 4. We will briefly summarize the main results obtained in the classical case. Thus, the Hawking temperature $T_0$, the Bekenstein-Hawking entropy $S_0$ and the heat capacity $C_0$ are computed to be \cite{BHreview,Biro:2017flp}:
\begin{align}
T_0 &= \frac{|f_0'(r_0)|}{4 \pi} = \frac{1}{4 \pi} 
\Bigg| \  \frac{3 \mu}{r_0^2} \ \Bigg|
 \propto \ \mu ^{1/3} ,
 \\
 S_0 & = \frac{\mathcal{A}_H}{4 G_0} = \frac{4 \pi^2 r_0^2}{4 G_0} 
\ \propto \ \mu^{2/3} ,
\\
C_0 &= T_0 \ \frac{\partial S_0}{\partial T_0} \ \Bigg|_{r_0} = -S_0 .
\end{align}
{\bf respectively, with $\mathcal{A}_H$ being the horizon area.}

%%%%%%%%%%%%%%%%%%%%%%%%%%%%%%%%%%%%%%%%%%%%%%%%%%%%%%%%%%
\subsection{Rotating solutions in scale-dependent gravity}
%%%%%%%%%%%%%%%%%%%%%%%%%%%%%%%%%%%%%%%%%%%%%%%%%%%%%%%%%%

We now apply the formalism presented in Section 2 to obtain the slowly rotating solution with flat horizon structure in four-dimensional scale-dependent gravity.  Let us remark in passing that the classical slowly rotating Kerr solution \cite{kerr} cannot be recovered, since the geometry of the Kerr BH is characterized by spherical horizon structure. Moreover, the three-dimensional full rotating Ba{\~n}ados-Zanelli-Teitelboim solution \cite{btzRot} in scale-dependent gravity has been obtained in \cite{SD5}. In four-dimensions, however, the complexity of the full field equations unfortunately does not allow for an exact treatment. Therefore, in the following we shall only focus on the limit of slow rotation.

\smallskip

We recall that the for the planar BH solution without rotation given by
\begin{equation}
ds^2 = -f(r) dt^2 + f(r)^{-1} dr^2 + \left(\frac{r}{L}\right)^2 (dx^2 + dy^2) ,
\end{equation}
the impact of scale-dependent gravity on its properties has been studied in \cite{SD8}. In particular, the modified lapse function and the running CC were found to be \cite{SD8}
\begin{eqnarray}
f(r) & = &  f_0(r) + 6 M L \epsilon Y(r) ,   \\
Y(r) & = & 1-2 \epsilon r + 2 (\epsilon r)^2 \ln\left( 1+\frac{1}{\epsilon r} \right) ,
\end{eqnarray}
and 
\begin{equation}
\Lambda(r) = \Lambda_0 + \epsilon \: \left( \frac{L}{ r (1+\epsilon r)^2} \right) \: \lambda(r) ,
\end{equation}
\begin{align}
\begin{split}
\lambda(r)  =  & 
\frac{\Lambda_0  r^2 (1+\epsilon r)}{L} + 6 M \epsilon \: [1+12 \epsilon r (1+\epsilon r)] - 
\\
&
36 M r \epsilon^2 (1+\epsilon r) (1+2 \epsilon r) \ln\left( 1+\frac{1}{\epsilon r} \right) ,
\end{split}
\end{align}
respectively, where the sub index 0 denotes the classical quantities, $\epsilon$ is the running parameter that measures the deviation from the classical solution.

\smallskip

We now proceed to obtain the scale-dependent version of the rotating solution in the slow rotation limit.
%
%%%%%%%%%%%%%%%%%%%%%%%%%%%%%%%%%%%%%%%%%%%%%%%%5
%
Based on the line element of the classical slow rotating solution, we here too make the following ansatz:
\begin{equation}
ds^2 = -f(r) dt^2 + f(r)^{-1} dr^2 + r^2 d \phi^2 + \frac{r^2}{L^2} dz^2 - 2 \omega n(r) \: dt d \phi ,
\end{equation}
where $f(r)$ and $n(r)$ are two unknown functions to be determined by the effective Einstein's field equations. Furthermore, since the angular velocity $\omega$ controls the rotation speed, in the slow rotation limit we shall keep terms linear in $\omega$ only. 
In addition, the specific form of Newton's coupling can be obtained directly, after eliminating $\Lambda(r)$, by solving the effective Einstein's field equations. Thus, we first take advantage of the $tt$ component of the effective field equations to obtain $G(r)$. Second, we compute the two metric functions, $f(r),n(r)$, which can be analytically obtained from the reduced system of differential equations. To conclude, we compute the running cosmological coupling $\Lambda(r)$. Thus, the full solution is found to be
\begin{align}
G(r) & =\frac{G_0}{1 +\epsilon r} ,
\\
f(r) & = f_0(r) + \frac{3}{2} \mu  \epsilon Y(r) ,
\\
n(r) & = n_0(r) - \frac{3}{2} \mu  L^2 \epsilon  Y(r),
\\
\begin{split}
\Lambda(r) & = \Lambda_0 + 
\frac{ \Lambda_0  \epsilon r }{(1 + \epsilon r)}
\Bigg[
1 - \frac{(1 + 12 r \epsilon  (1 + \epsilon r)) \left(\mu  L^2 \epsilon \right)}{2 r^2 (1 + \epsilon r)} \ +
\\
&
\hspace{3.1cm}
\frac{(1 + 2 r \epsilon) \left(3 \mu  L^2 \epsilon ^2\right)}{r} \ln \left(1 + \frac{1}{r \epsilon } \right)
\Bigg],
\end{split}
\end{align}
where for convenience we introduce an auxiliary function, $Y(r)$, which is defined to be:
\begin{equation}
Y(r) \equiv 1-2 \epsilon r + 2 (\epsilon r)^2 \ln\left( 1+\frac{1}{\epsilon r}\right).
\end{equation}
At this point, a couple of comments are in order. First, the integration constants have been chosen conveniently to obtain the classical solution when the scale-dependent parameter is taken to be zero. It can be observed below:
\begin{align}
\lim_{\epsilon \rightarrow 0} G(r) &= G_0 \equiv 1  ,
\\
\lim_{\epsilon \rightarrow 0} f(r) &= f_0(r) = -\frac{\mu}{r} + \frac{r^2}{L^2} ,
\\
\lim_{\epsilon \rightarrow 0} n(r) &= n_0(r) \equiv \frac{\mu L^2}{r} ,
\\
\lim_{\epsilon \rightarrow 0} \Lambda(r) &= \Lambda_0 \equiv -\frac{3}{L^2}.
\end{align}
Moreover, when the running parameter is small enough, we can take a series for $ \epsilon $ parameter. Thus, our new solution is contrasted with the classical one. In this respect, we also observe the leading corrections, which are summarized as follows:
\begin{align}
G(r) &\approx G_0 (1 - \epsilon r) + \mathcal{O}(\epsilon^2) ,
\\
f(r) &\approx f_0(r) + \frac{3}{2} \mu \epsilon + \mathcal{O}(\epsilon^2) ,
\\
n(r) &\approx n_0(r)\left(1 - \frac{3}{2} \epsilon  r\right) + \mathcal{O}(\epsilon^2) ,
\\
\Lambda(r) &\approx \Lambda_0 (1 + \epsilon r) + \mathcal{O}(\epsilon^2).
\end{align}

\smallskip 

To summarize, scale-dependent solutions are obtained taking into account two ingredients: 
i) Einstein's effective field equations, and ii) the link between the renormalization scale $k$ and the radial coordinate $r$. The latter is a reasonable assumption, and it has been used in similar problems, such as improved BH solutions. With the above in mind, the classical couplings are treated as scale-dependent ones, and we need to solve the improved Einstein's field equations.

\smallskip

The general process to obtain the solutions in this case may be summarized as follows: 
i) given that $\Lambda(r)$ appears linearly in the field equations, we may eliminate it
temporarily combining, ii) After that, the reduced $tt$ component system of differential equations allows us to obtain $G(r)$ directly, ignoring the rest of unknown functions $f(r), n(r)$. iii) Then, plugging in $G(r)$ and $\Lambda(r)$, we use the $\phi \phi$ effective field equation to obtain the lapse function.
iv) Plugging $\Lambda(r), G(r), f(r)$ into the non-diagonal part of Einstein's field equations we compute $n(r)$.
v) Finally, we plug $G(r), f(r), n(r)$ into the original equations to obtain the explicit form of $\Lambda(r)$.

%%%%%%%%%%%%%%%%%%%%%%%%%%%%FIGURES%%%%%%%%%%%%%%%%%%%%%%%%%%%%%%%%%%%%%

\begin{figure*}[ht!]
\centering
\includegraphics[width=0.48\textwidth]{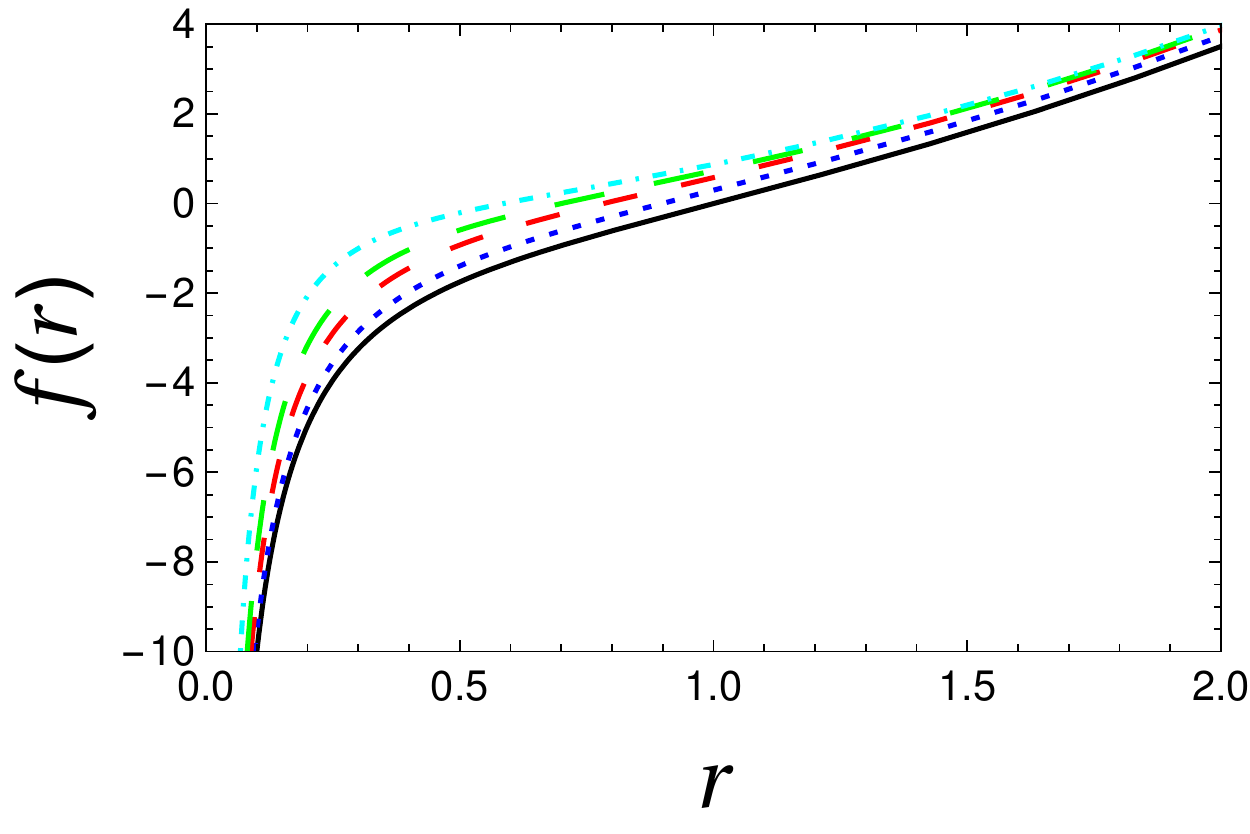}   \ \
\includegraphics[width=0.48\textwidth]{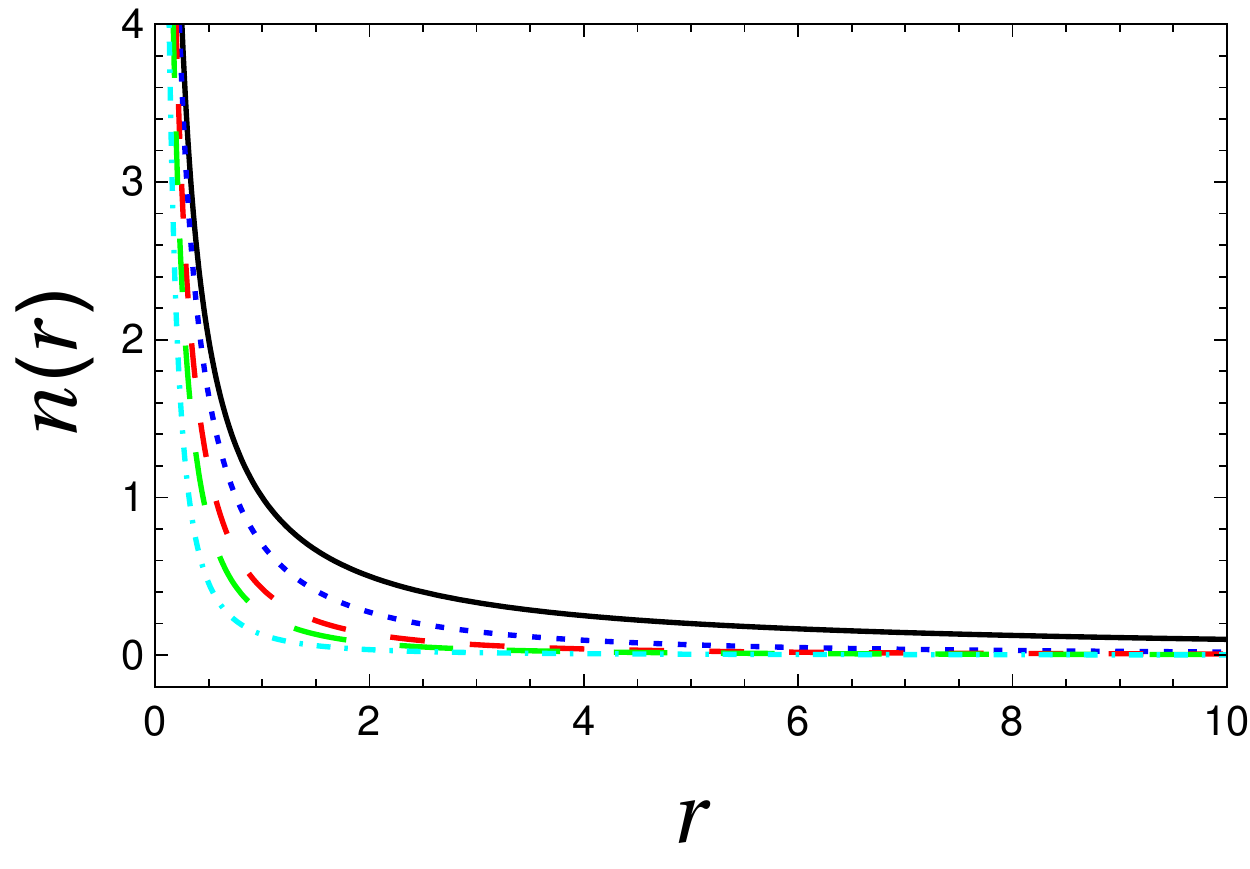}  
\\
\includegraphics[width=0.48\textwidth]{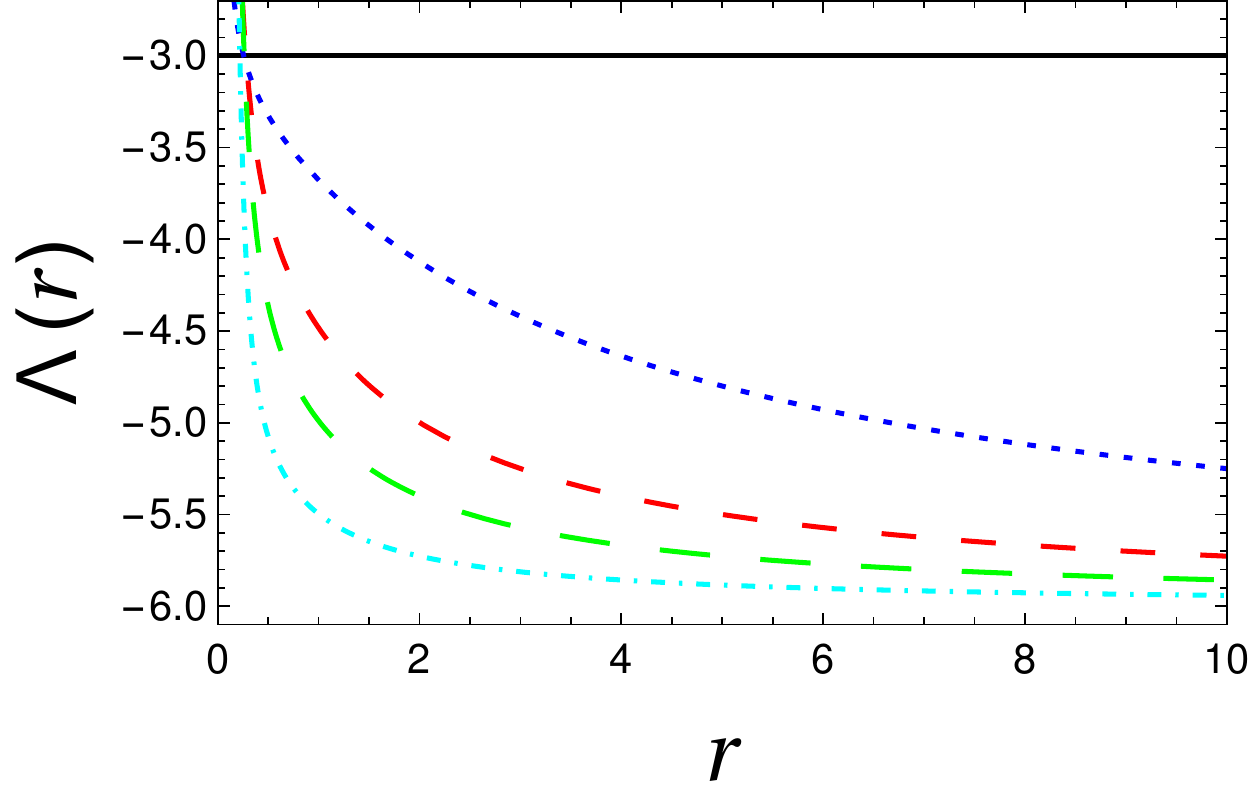}   \ \
\includegraphics[width=0.48\textwidth]{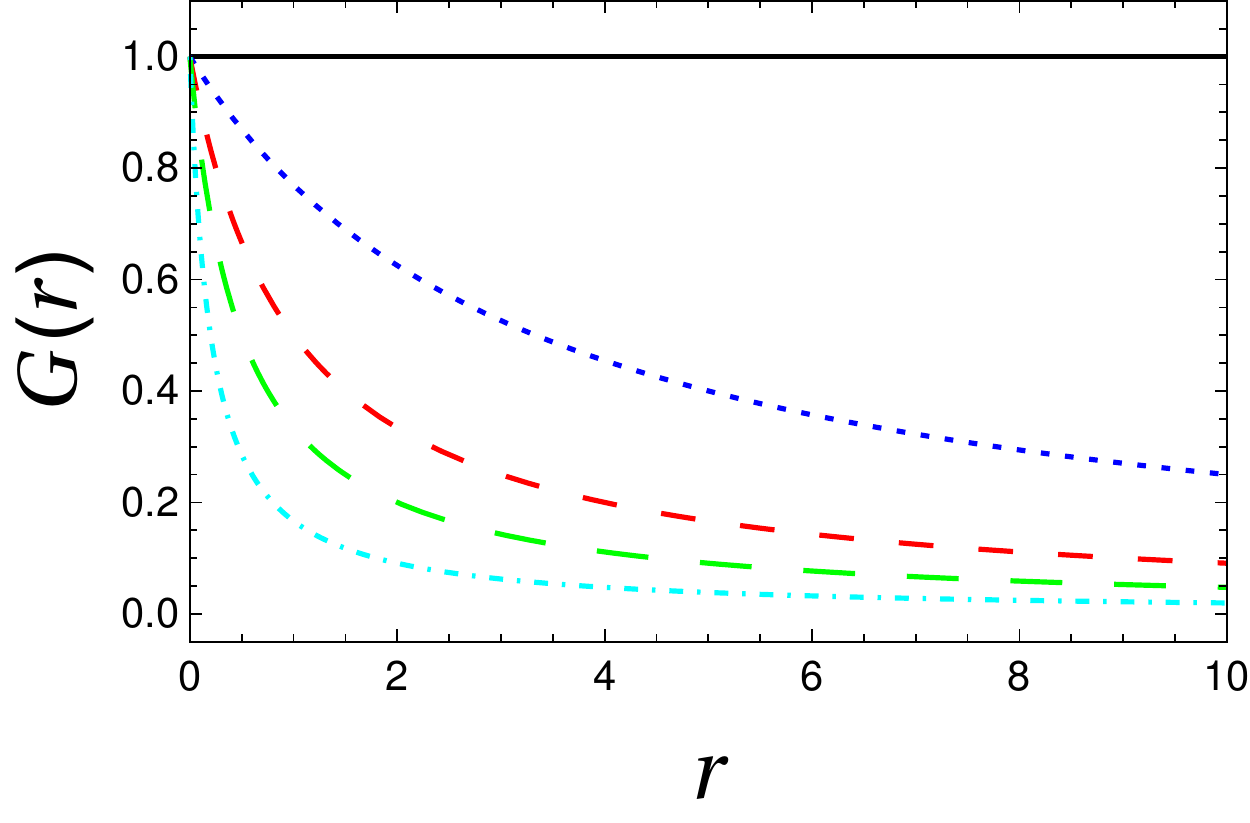}  
%}
\caption{ 
Metric functions $f(r),n(r)$ as well as running cosmological constant, $\Lambda(r)$, and Newton's constant, $G(r)$, setting $\mu=1$ and $L=1$.
{\bf Top left panel:} Lapse function $f(r)$ versus $r$ for different values of the running parameter $\epsilon$. 
{\bf Top right panel:} Shift function $n(r)$ versus $r$ for different values of the running parameter $\epsilon$. 
{\bf Bottom left panel:} Running cosmological constant $\Lambda(r)$ versus $r$ for different values of the running parameter $\epsilon$. 
{\bf Bottom right panel:} Running Newton's constant $G(r)$ versus $r$ for different values of the running parameter $\epsilon$.
Shown are: 
i)   $\epsilon =0$ (solid black line)
ii)  $\epsilon =0.3$ (dotted blue line)
iii) $\epsilon =1$ (short-dashed red line)
iv)  $\epsilon =2$ (long dashed green line)
v)   $\epsilon =5$ (dotted-dashed cyan line). Notice that $\epsilon=0$ corresponds to the classical quantities, which are also shown for comparison reasons.
}
\label{fig:1}
\end{figure*}

%%%%%%%%%%%%%%%%%%%%%%%%%%%%%%%%%%%%%%%%%%%%%%%%%%%%%

\begin{figure*}[ht!]
\centering
\includegraphics[width=0.48\textwidth]{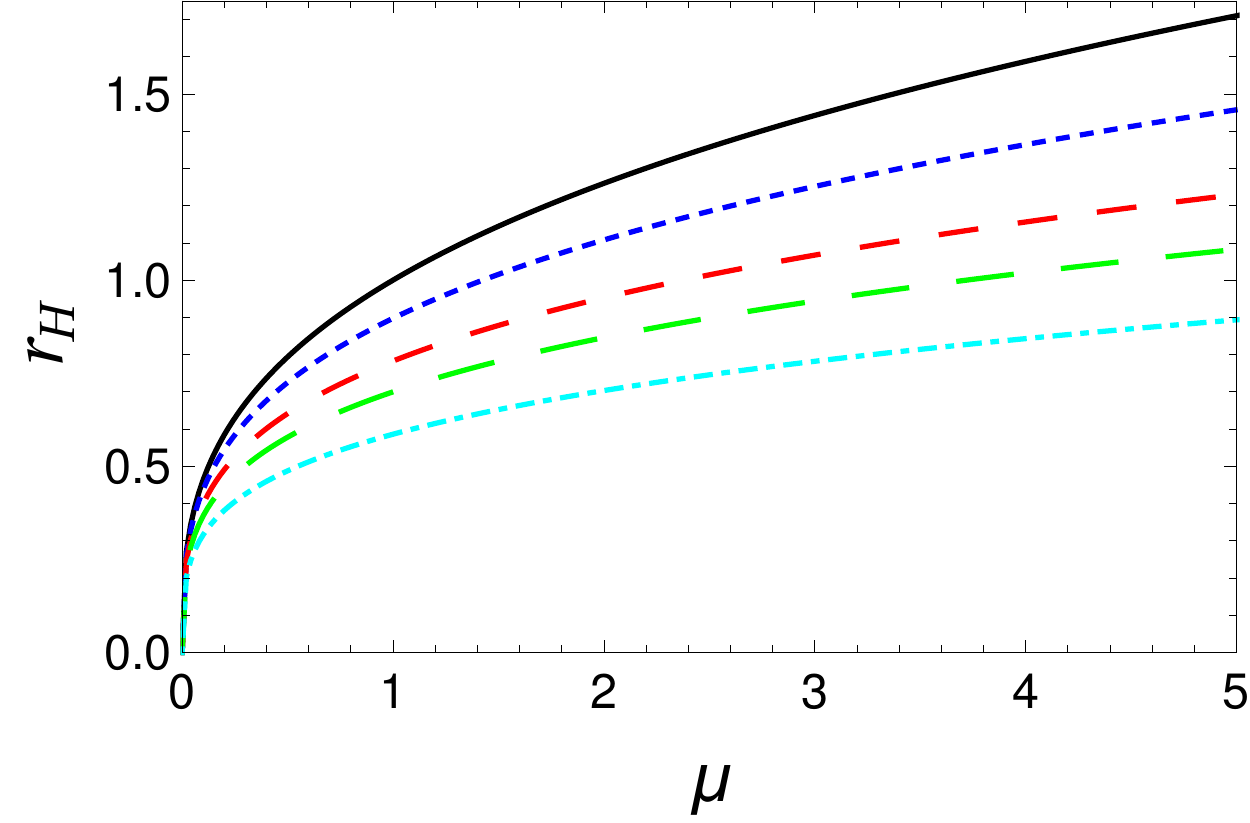}   \ \
\includegraphics[width=0.48\textwidth]{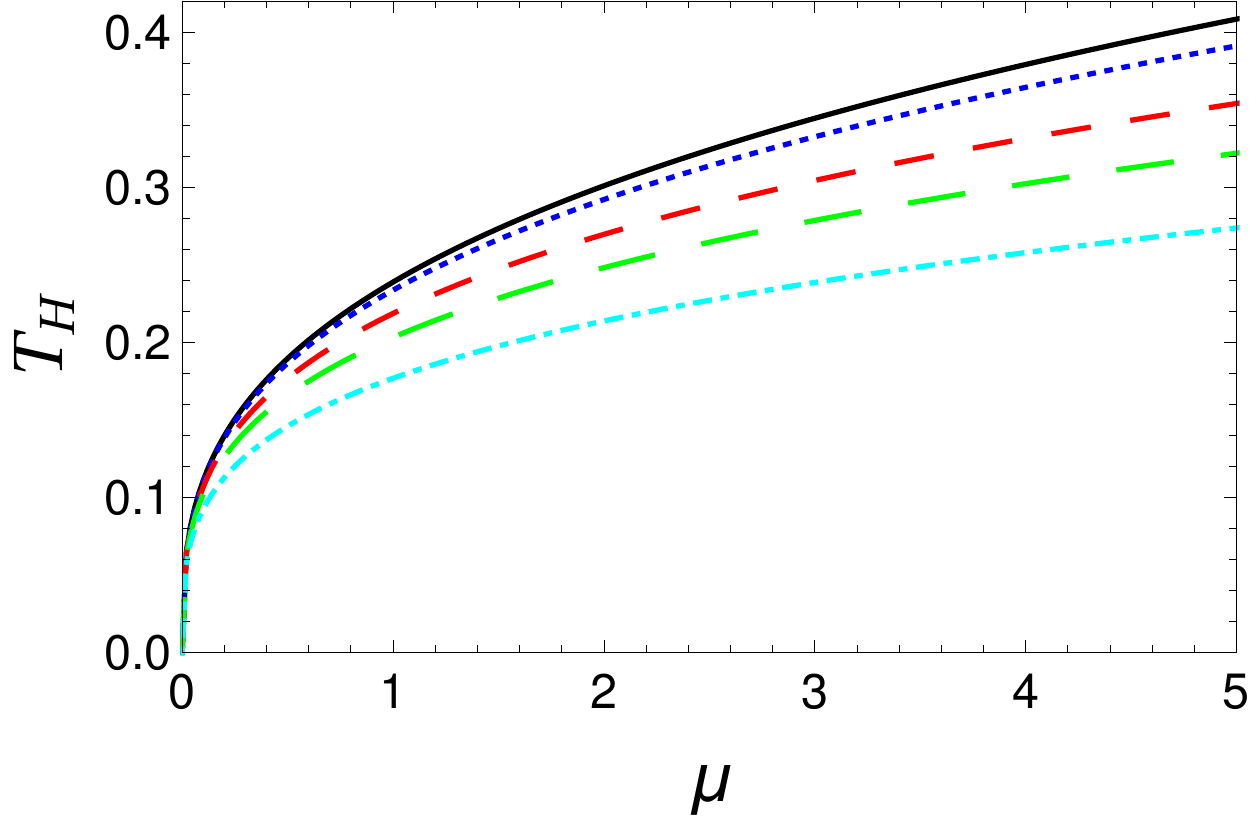}  
\\
\includegraphics[width=0.48\textwidth]{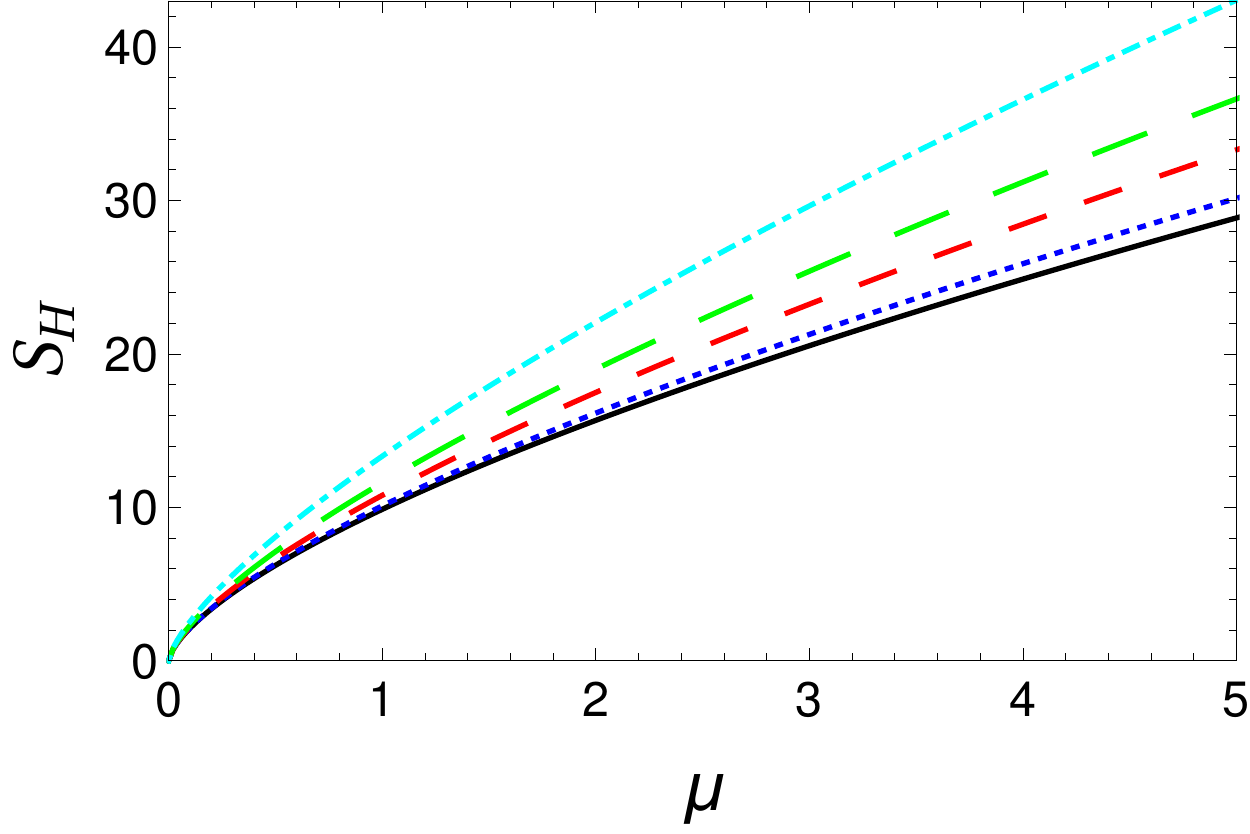}   \ \
\includegraphics[width=0.48\textwidth]{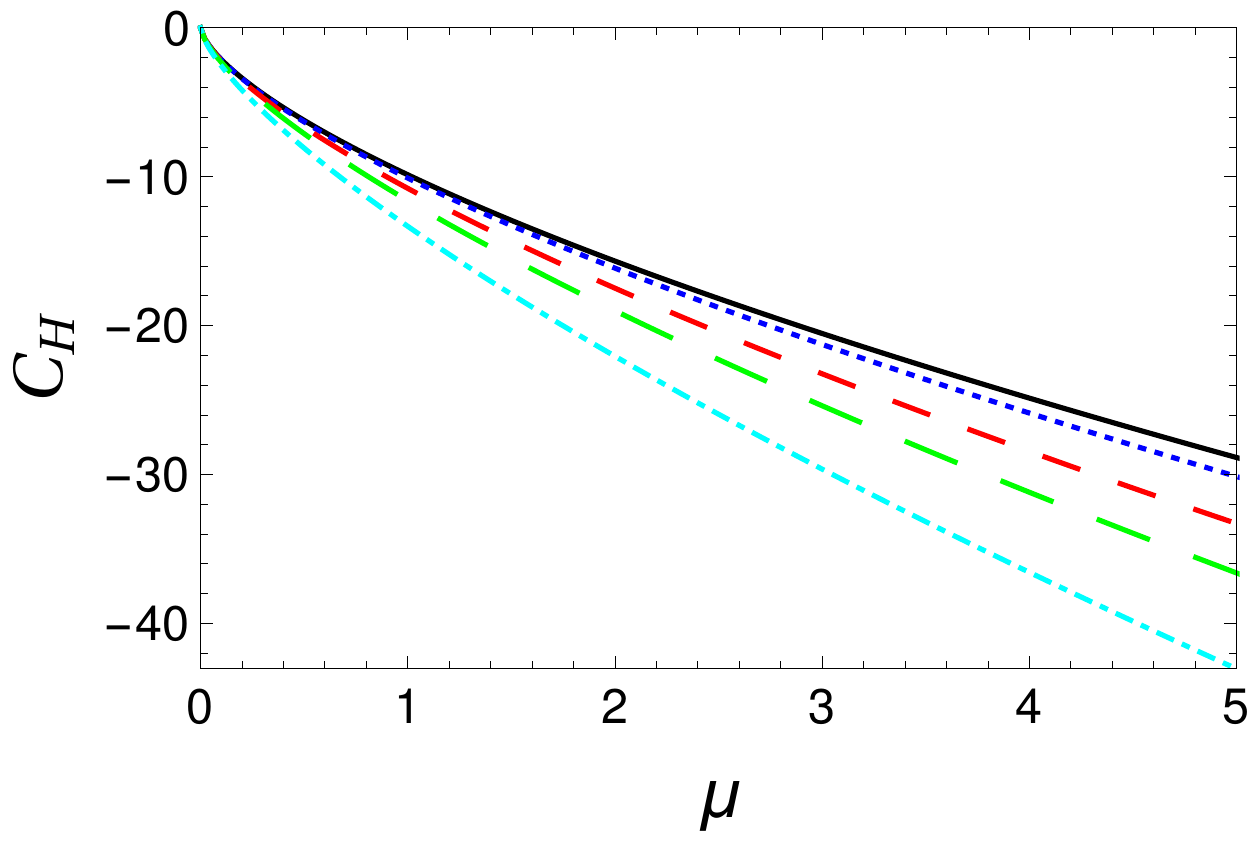}  
%}
\caption{
Event horizon and thermodynamic quantities setting $L=1$.
{\bf Top left panel:} BH horizon $r_H$ versus $\mu$ for different values of the running parameter $\epsilon$. 
{\bf Top right panel:} Hawking temperature $T_H$ versus $\mu$ for different values of the running parameter $\epsilon$. 
{\bf Bottom left panel:} Bekenstein-Hawking entropy $S_H$ versus $\mu$ for different values of the running parameter $\epsilon$. 
{\bf Bottom right panel:} Specific heat $C_H$ versus $\mu$ for different values of the running parameter $\epsilon$.
Shown are: 
i)   $\epsilon =0$ (solid black line)
ii)  $\epsilon =0.3$ (dotted blue line)
iii) $\epsilon =1$ (short-dashed red line)
iv)  $\epsilon =2$ (long dashed green line)
v)   $\epsilon =5$ (dotted-dashed cyan line).  Notice that $\epsilon=0$ corresponds to the classical quantities, which are also shown for comparison reasons.
}
\label{fig:2}
\end{figure*}

%%%%%%%%%%%%%%%%%%%%%%%%%%%FIGURES%%%%%%%%%%%%%%%%%%%%%%%%%%%%

%%%%%%%%%%%%%%%%%%%%%%%%%%%%%%%%%%%%%%%%
\section{Invariants and thermodynamics}
%%%%%%%%%%%%%%%%%%%%%%%%%%%%%%%%%%%%%%%%

It is always useful to investigate BH thermodynamics as well as some of the invariants of the theory. This analysis in principle could reveal new non-physical singularities, which should be treated in some detail. Also, thermodynamic properties allow us to get some insight into the underlying theory. It is known that the well-known solutions of General Relativity are characterized by a singularity at the origin, $r \rightarrow 0$. The singularity is hidden by an event horizon, and therefore it has no effect on the outside region, where Physics is well-behaved. The existence of singularities, however, indicates the breakdown of General Relativity, and understanding of the final stages of gravitational collapse is not possible when singularities are present.

\smallskip

Before we start, it is essential to point out that given the complexity of the lapse function, it is not possible to obtain an expression for the horizon in a closed form. The horizon is computed solving the algebraic equation $f(r_H) = 0$, with $r_H$ being the scale-dependent BH horizon. To get some intuition, we may obtain approximate expressions assuming a small running parameter, although the figures have been produced using the full expressions, and therefore $\epsilon$ is not required to be small. We thus can expand $f(r)$ in powers of $\epsilon$, and use in the following an approximate expression for the lapse function
\begin{align}
f(r) &\approx f_0(r) + \frac{3}{2} \mu \epsilon - 3 \mu \epsilon ^2 r.
\end{align}
The expression for the BH horizon is then found to be
\begin{align}
r_H = r_0 \Bigg(1 - \frac{1}{2}(\epsilon r_0) + (\epsilon r_0)^2  + \mathcal{O}(\epsilon^3)\Bigg).
\end{align}
We see that the new horizon is smaller than the one found in the classical solution. Besides, as can be read off from Fig.~\eqref{fig:2}, the corrections appear for large values of the mass parameter $\mu$. After that, we will move to the computation of the black hole invariants as well as the basic thermodynamic quantities, such as temperature, entropy and heat capacity.

\subsection{Invariants}

As already mentioned before, a full analysis of the invariants is also relevant due to the fact that it reveals the presence of potentially new singularities. Here we shall compute two of them, i.e., i) the Ricci scalar $\mathcal{R}$ and ii) the Kretschmann scalar $\mathcal{K}$.

\subsubsection{Ricci scalar $\mathcal{R}$}

In the slow rotation limit, the value of the Ricci scalar coincides with the one corresponding to the non-rotating case, due to the fact that the contribution of the rotation speed is proportional to $\omega^2$, which is of higher order and thus neglected. In differential geometry, the Ricci scalar $\mathcal{R}$ is computed starting from the metric tensor and computing the Christoffel symbols $\Gamma^{\mu}_{\rho \sigma}$ and the Ricci tensor $R_{\mu \nu}$ first as follows \cite{Riotto}
\begin{equation}
\Gamma^\mu_{\rho \sigma} = \frac{1}{2} g^{\mu \lambda} \left( \frac{\partial g_{\lambda \rho}}{\partial x^\sigma} + \frac{\partial g_{\lambda \sigma}}{\partial x^\rho} - \frac{\partial g_{\rho \sigma}}{\partial x^\lambda} \right) ,
\end{equation}
\begin{equation}
R_{\mu \nu} = \partial_\lambda \Gamma^{\lambda}_{\mu \nu} - \partial_{\mu} \Gamma^{\lambda}_{\lambda \nu} + \Gamma^{\lambda}_{\lambda \sigma} \Gamma^{\sigma}_{\mu \nu} - \Gamma^{\lambda}_{\mu \sigma} \Gamma^{\sigma}_{\lambda \nu} ,
\end{equation}
\begin{equation}
\mathcal{R} = g^{\mu \nu} R_{\mu \nu} ,
\end{equation}
 Therefore, the Ricci scalar is finally computed to be
\begin{align}
\mathcal{R} &= -f''(r) -\frac{4 f'(r)}{r}-\frac{2 f(r)}{r^2},
\end{align}
where $f(r)$ is the new scale-dependent lapse function. We then substitute its expression to obtain:
\begin{align} \label{R}
\mathcal{R} &= -\frac{12}{L^2} +   3 \mu  \epsilon \frac{(1 + 2 \epsilon r) (6 \epsilon r ( 1 +  \epsilon r) - 1 )}{r^2 (1 + \epsilon r)^2} -36 \mu  \epsilon ^3 \ln \left(1 + \frac{1}{\epsilon r} \right),
\end{align}
where the classical value is a constant, which is computed to be
\begin{equation}
\mathcal{R}_0 = -\frac{12}{L^2} ,
\end{equation}
and which is recovered when $\epsilon$ is set to zero. Finally, the second term in Eq.~\eqref{R} exhibits a new singularity due to quantum effects. The same holds for the logarithmic term, which blows up when the radial coordinate goes to zero. Given that we are interested in small deviations from the classical solution, we expand around $\epsilon=0$ once more to obtain
\begin{align}
\mathcal{R} &= \mathcal{R}_0  -\frac{3 \mu  \epsilon }{r^2}  +  \frac{18 \mu  \epsilon ^2}{r} + \mathcal{O}(\epsilon^3).
\end{align}
Therefore, we confirm that the Ricci scalar has a  single singularity at the origin.

\subsubsection{Kretschmann scalar $\mathcal{K}$}

We shall now investigate how the Kretschmann scalar is affected when the running of the coupling constants of the theory is considered. Once more, for the slowly rotating solutions the Kretschmann scalar, which is defined to be 
\begin{equation}
\mathcal{K} \equiv R_{abcd} R^{abcd} ,
\end{equation}
with $R_{abcd}$ being the Riemann tensor, takes the simple form
\begin{align}
\mathcal{K} &= f''(r)^2 + \frac{4 f'(r)^2}{r^2} - \frac{4 f(r)^2}{r^4}.
\end{align}
In this case, the expression for $\mathcal{K}$ becomes quite complicated, which is why we will only focus on its approximated expression. Thus, when $\epsilon$ is small, $\mathcal{K}$ acquires the approximate form
\begin{align}
\mathcal{K} &= \mathcal{K}_0 + (\epsilon r)  \left(\frac{12 \mu }{L^2 r^3} - \frac{12 \mu ^2}{r^6}\right)  +  \mathcal{O}(\epsilon^2),
\end{align}
where the classical value is found to be
\begin{align}
\mathcal{K}_0 &\equiv \frac{24}{L^4}  + \frac{12 \mu ^2}{r^6}.
\end{align}
We see a single singularity at the origin, $r \rightarrow 0$, both in the classical theory and in scale-dependent gravity. Therefore, scale-dependent gravity is not able to eliminate the singularity of classical theory. In the former additional terms that blow up at the origin are present, although as $r \rightarrow 0$ the classical contribution is the dominant one. This confirms what we have already seen in Fig.~\ref{fig:1}  (panel for $f(r)$), where there are no deviations from the classical theory at $r=0$, and where, as already mentioned before, variations only occur at intermediate scales. We also observe that the scale-dependent effect slightly increases the invariant since the running parameter is always taken to be small, to maintain the deviations (from its classical value) under control.

\subsection{Thermodynamics}

In the following we will discuss the basic thermodynamic properties to get some insight into the physics behind the scale-dependent black hole solutions.

\smallskip

Before we start, we should point out how BH thermodynamics is deformed when scale-dependent gravity is considered. We will focus on three concrete thermodynamic quantities, namely: i) Hawking temperature, ii) Bekenstein-Hawking entropy, and iii) Heat capacity.
The first quantity, i.e., the Hawking temperature, can be computed by standard means for $T_H$, the only difference being that the metric potentials are deformed. Therefore, the modification due to the formalism does not change the standard formula for $T_H $ valid in GR. To be more precise, we can recognize that $T_0$ and $T_H$ share the same functional form by noticing that Newton's coupling is promoted from a constant, $G_0=1$, to a r-varying $G(r)=1/(1 + \epsilon r)$ function.
Second, the Bekenstein-Hawking entropy within scale-dependent gravity can be obtained from the 
Brans-Dicke theory. In particular, similarly to the Hawking temperature, replacing $G_0$ by $G(r)$, 
we obtain an improved relation for $S_H$. 
Finally, as we will show, the heat capacity once more may be computed using the classical relation. 
Thus, roughly speaking, $C_H$ is deformed due to $G_0 \rightarrow G(r)$ as before. All three effects are clearly observed in Fig.~\eqref{fig:2}.

\subsubsection{Hawking temperature}

We will first introduce the Hawking temperature of the scale-dependent black hole solution in four-dimensional space-time. Following the same procedure as in the classical solution, we compute $T_H$ as follows  (see \cite{Cai:1994np}) :
\begin{align} %\nonumber
T_H &= \frac{1}{4 \pi} 
\Bigg| 
\lim_{r \rightarrow r_H} \frac{\partial_r g_{tt}}{\sqrt{-g_{tt}g_{rr}}} \Bigg|
 =
 \frac{1}{4 \pi} \Bigg| \ \frac{3 \mu}{r_H^2 (1 + \epsilon r_H)} \ \Bigg|.
\end{align}
Clearly, when $\epsilon$ tends to zero, the classical solution is recovered.
We should observe how the scale-dependent formalism introduces deviations from its classical counterpart. Given the last expression, we see that the temperature decreases in comparison with the classical case. This becomes clearer, taking an expansion for small values of $\epsilon$, and rewriting it in terms of the classical horizon, namely:
\begin{align}
T_H &= T_0 \Bigg|  \left(1 - \frac{3}{4} (\epsilon r_0)^2 \right) + \mathcal{O}(\epsilon^3)  \ \Bigg|.
\end{align}
Fig.~\eqref{fig:1} confirms that when the mass term increases, the scale-dependent temperature is lower than its classical value. It coincides with the classical solution for small values of $\mu$ (both in the classical and the scale-dependent solution).

\subsubsection{Bekenstein-Hawking entropy}

Another thermodynamic property to be analyzed is the well-known Bekenstein-Hawking entropy \cite{Gibbons:1976ue}. The approach followed here may be viewed as a particular case of a scalar-tensor theory of gravity, and therefore the corresponding extended formula for this type of theories is given by \cite{Kang:1996rj} 
\begin{align}
S &= \frac{1}{4} \oint  {\mathrm {d}^2}x \frac{\sqrt{h}}{G(x)},
\end{align}
where $h_{ij}$ is the induced metric at the horizon. Taking advantage of the symmetry as well as the fact that $G(x) = G(r_H )$ is constant along the
horizon, the above integral takes the form \cite{SD1,SD0}
\begin{align}\label{eqSs}
S_H &=\frac{\mathcal{A}_H}{4G(r_H)} = S_0(r_H)(1 + \epsilon r_H).
\end{align}
Notice that the entropy is larger than the one corresponding to the classical solution when the mass parameter increases (see Fig.~\eqref{fig:2} for details). Also, in contrast to the standard solution, where $S_0$ is proportional to the horizon area, our expression (based on the Brans-Dicke approach) mimics an ``area $ \times $ radio" law. As it should be, we also recover the classical solution when the scale-dependent parameter is set to zero. Similar to previous quantities, the Bekenstein-Hawking entropy largely differs from the classical solution for large values of the scale-dependent parameter. It coincides with the classical values when $\epsilon$ is taken to be zero.

%%%%%%%%%%%%%%%%%%%%%%%%%%%%%%
\subsubsection{Heat capacity}
%%%%%%%%%%%%%%%%%%%%%%%%%%%%%%

The heat capacity is computed making use of the usual relation
\begin{align}
C_H &= T_H \ \frac{\partial S_H}{\partial T_H} \ \Bigg|_{r_H} = -S_H,
\end{align}
where the numerical solution is also shown in Fig.~\eqref{fig:2}.  It is essential to point out that, similarly to the entropy, the latter is an exact result, and it tends to the classical solution when $\epsilon \rightarrow 0$.

\smallskip

What is more, given that $S \sim \mu^{2/3}$  and positive, the above equation implies that the heat capacity is negative \cite{Biro:2017flp}.  This is in agreement with the well-known fact that in all bound systems with positive kinetic energy and total negative energy, an increase of the temperature appears, and the total energy will decrease, producing a negative heat capacity. In light of the previous comments, thermal equilibrium between a negative specific heat system and a positive one is not possible, which is the reason why BHs in this sense seem to be thermally unstable. The same holds for the scale-dependent solutions, where the inclusion of quantum features does not substantially alter the underlying behaviour. Therefore, the solution in scale-dependent gravity is still unstable, as the classical one.

\smallskip

Before we conclude our work, let us briefly comment on future work. Nowadays, gravitational wave astronomy \cite{valeria} and quasinormal modes of black holes \cite{review1,review2,review3}
is a very active field. Moreover, after the first image of the shadow of a supermassive BH \cite{L1,L4,L5,L6}, studying the shadows that rotating black holes in several different contexts can also cast become an exciting field. Therefore, we feel it would be interesting to compute the quasinormal modes and the shadow of the slowly rotating scale-dependent solution obtained here. We hope to be able to address those issues in forthcoming publications.

%%%%%%%%%%%%%%%%%%%%%%%%
\section{Conclusions}
%%%%%%%%%%%%%%%%%%%%%%%%

In summary, in this work, we have studied some of the properties of four-dimensional slowly rotating BHs with a flat horizon structure in the scale-dependence scenario. Starting from the average effective action, we have computed the corresponding effective Einstein's field equations, and we have obtained the functions involved. In the slow-rotating limit, the combination $\omega n(r)$ encodes the rotation of the black hole, with $ \omega $ being the angular velocity. As can be observed in Fig.~\eqref{fig:1}, the function $n(r)$ mimics the classical behaviour for large and small values of the radial coordinate (the same occurs with the lapse function). Thus, the deviations from the classical solution are significant only in the intermediate region. 

\smallskip

Note that contrary to other scale-dependent solutions, no energy conditions have been used here. We also have investigated the basic thermodynamic properties of this model, observing that they are slightly modified after the inclusion of the scale-dependent couplings. Finally, we have investigated the invariants of the theory, according to which both in the classical theory and in scale-dependent gravity, there is a single singularity at the origin, $r \rightarrow 0$. Our study reveals small deviations in the IR region, consistent with predictions based on asymptotically safe gravity.

%%%%%%%%%%%%%%%%%%%%%%%%%%%%%%%%%%%%%%%%%%%%%%%%%%%%%%%%%%%%%%%%%%%%%%%%%%%

\section*{Acknowlegements}

We wish to thank the anonymous reviewer for a careful reading of the manuscript, 
as well as for numerous useful comments and suggestions. 
The author A.~R. acknowledges DI-VRIEA for financial support through Proyecto 
Postdoctorado 2019 VRIEA-PUCV. The author G.~P. thanks the Funda\c c\~ao para a 
Ci\^encia e Tecnologia (FCT), Portugal, for the financial support to the Center for 
Astrophysics and Gravitation-CENTRA, Instituto Superior T\'ecnico, Universidade de Lisboa, 
through the Project No.~UIDB/00099/2020.

%%%%%%%%%%%%%%%%%%%%%%%%%%%%%%%%%%%%%%%%%%%%%%%%%%%%%%%%%%%%%%%%%%%%%%%%%%%

%\newpage


\begin{thebibliography}{99}
%
\bibitem{GR}
A.~Einstein,
%``The Foundation of the General Theory of Relativity,''
Annalen Phys. \textbf{49} (1916) no.7, 769-822
%doi:10.1002/andp.200590044

\bibitem{tests1}
S.~G.~Turyshev,
%``Experimental Tests of General Relativity,''
Ann. Rev. Nucl. Part. Sci. \textbf{58} (2008), 207-248
%doi:10.1146/annurev.nucl.58.020807.111839
[arXiv:0806.1731 [gr-qc]].

\bibitem{tests2}
C.~M.~Will,
%``The Confrontation between General Relativity and Experiment,''
Living Rev. Rel. \textbf{17} (2014), 4
%doi:10.12942/lrr-2014-4
[arXiv:1403.7377 [gr-qc]].
  
\bibitem{tests3}
E.~Asmodelle,
%``Tests of General Relativity: A Review,''
[arXiv:1705.04397 [gr-qc]].

\bibitem{Donoghue:1994dn}
J.~F.~Donoghue,
%``General relativity as an effective field theory: The leading quantum corrections,''
Phys. Rev. D \textbf{50} (1994), 3874-3888
%doi:10.1103/PhysRevD.50.3874
[arXiv:gr-qc/9405057 [gr-qc]].
    
\bibitem{QG1}
T.~Jacobson,
%``Thermodynamics of space-time: The Einstein equation of state,''
Phys. Rev. Lett. \textbf{75} (1995), 1260-1263
%doi:10.1103/PhysRevLett.75.1260
[arXiv:gr-qc/9504004 [gr-qc]].  
  
\bibitem{QG2}
A.~Connes,
%``Gravity coupled with matter and foundation of noncommutative geometry,''
Commun. Math. Phys. \textbf{182} (1996), 155-176
%doi:10.1007/BF02506388
[arXiv:hep-th/9603053 [hep-th]].

\bibitem{QG3}
M.~Reuter,
%``Nonperturbative evolution equation for quantum gravity,''
Phys. Rev. D \textbf{57} (1998), 971-985
%doi:10.1103/PhysRevD.57.971
[arXiv:hep-th/9605030 [hep-th]].

\bibitem{QG4}
C.~Rovelli,
%``Loop quantum gravity,''
Living Rev. Rel. \textbf{1} (1998), 1
%doi:10.12942/lrr-1998-1
[arXiv:gr-qc/9710008 [gr-qc]].

\bibitem{QG5}
R.~Gambini and J.~Pullin,
%``Consistent discretization and loop quantum geometry,''
Phys. Rev. Lett. \textbf{94} (2005), 101302
%doi:10.1103/PhysRevLett.94.101302
[arXiv:gr-qc/0409057 [gr-qc]].

\bibitem{QG6}
A.~Ashtekar,
%``Gravity and the quantum,''
New J. Phys. \textbf{7} (2005), 198
%doi:10.1088/1367-2630/7/1/198
[arXiv:gr-qc/0410054 [gr-qc]].

\bibitem{QG7}
P.~Nicolini,
%``Noncommutative Black Holes, The Final Appeal To Quantum Gravity: A Review,''
Int. J. Mod. Phys. A \textbf{24} (2009), 1229-1308
%doi:10.1142/S0217751X09043353
[arXiv:0807.1939 [hep-th]].

\bibitem{QG8}
P.~Horava,
%``Quantum Gravity at a Lifshitz Point,''
Phys. Rev. D \textbf{79} (2009), 084008
%doi:10.1103/PhysRevD.79.084008
[arXiv:0901.3775 [hep-th]].

\bibitem{QG9}
E.~P.~Verlinde,
%``On the Origin of Gravity and the Laws of Newton,''
JHEP \textbf{04} (2011), 029
%doi:10.1007/JHEP04(2011)029
[arXiv:1001.0785 [hep-th]].
  
\bibitem{miniBH1}
I.~Antoniadis, E.~Kiritsis and T.~N.~Tomaras,
%``A D-brminiBH1ane alternative to unification,''
Phys. Lett. B \textbf{486} (2000), 186-193
%doi:10.1016/S0370-2693(00)00733-4
[arXiv:hep-ph/0004214 [hep-ph]].  
  
  \bibitem{miniBH2}
A.~Mironov, A.~Morozov and T.~N.~Tomaras,
%``Can centauros or chirons be the first observations of evaporating mini black holes?,''
Int. J. Mod. Phys. A \textbf{24} (2009), 4097-4115
%doi:10.1142/S0217751X09044693
[arXiv:hep-ph/0311318 [hep-ph]].
  
\bibitem{miniBH3}
A.~Casanova and E.~Spallucci,
%``TeV mini black hole decay at future colliders,''
Class. Quant. Grav. \textbf{23} (2006), R45-R62
%doi:10.1088/0264-9381/23/3/R01
[arXiv:hep-ph/0512063 [hep-ph]].

\bibitem{L1}
K.~Akiyama \textit{et al.} [Event Horizon Telescope],
%``First M87 Event Horizon Telescope Results. I. The Shadow of the Supermassive Black Hole,''
Astrophys. J. \textbf{875} (2019) no.1, L1
%doi:10.3847/2041-8213/ab0ec7
[arXiv:1906.11238 [astro-ph.GA]].
  
  \bibitem{L2}
K.~Akiyama \textit{et al.} [Event Horizon Telescope],
%``First M87 Event Horizon Telescope Results. II. Array and Instrumentation,''
Astrophys. J. Lett. \textbf{875} (2019) no.1, L2
%doi:10.3847/2041-8213/ab0c96
[arXiv:1906.11239 [astro-ph.IM]].

\bibitem{L3}
K.~Akiyama \textit{et al.} [Event Horizon Telescope],
%``First M87 Event Horizon Telescope Results. III. Data Processing and Calibration,''
Astrophys. J. Lett. \textbf{875} (2019) no.1, L3
%doi:10.3847/2041-8213/ab0c57
[arXiv:1906.11240 [astro-ph.GA]].

\bibitem{L4}
K.~Akiyama \textit{et al.} [Event Horizon Telescope],
%``First M87 Event Horizon Telescope Results. IV. Imaging the Central Supermassive Black Hole,''
Astrophys. J. Lett. \textbf{875} (2019) no.1, L4
%doi:10.3847/2041-8213/ab0e85
[arXiv:1906.11241 [astro-ph.GA]].

\bibitem{L5}
K.~Akiyama \textit{et al.} [Event Horizon Telescope],
%``First M87 Event Horizon Telescope Results. V. Physical Origin of the Asymmetric Ring,''
Astrophys. J. Lett. \textbf{875} (2019) no.1, L5
%doi:10.3847/2041-8213/ab0f43
[arXiv:1906.11242 [astro-ph.GA]].

\bibitem{L6}
K.~Akiyama \textit{et al.} [Event Horizon Telescope],
%``First M87 Event Horizon Telescope Results. VI. The Shadow and Mass of the Central Black Hole,''
Astrophys. J. Lett. \textbf{875} (2019) no.1, L6
%doi:10.3847/2041-8213/ab1141
[arXiv:1906.11243 [astro-ph.GA]].

 \bibitem{ligo1}
B.~P.~Abbott \textit{et al.} [LIGO Scientific and Virgo],
%``Observation of Gravitational Waves from a Binary Black Hole Merger,''
Phys. Rev. Lett. \textbf{116} (2016) no.6, 061102
%doi:10.1103/PhysRevLett.116.061102
[arXiv:1602.03837 [gr-qc]].
  
\bibitem{ligo2}
B.~P.~Abbott \textit{et al.} [LIGO Scientific and Virgo],
%``GW151226: Observation of Gravitational Waves from a 22-Solar-Mass Binary Black Hole Coalescence,''
Phys. Rev. Lett. \textbf{116} (2016) no.24, 241103
%doi:10.1103/PhysRevLett.116.241103
[arXiv:1606.04855 [gr-qc]].

\bibitem{ligo3}
B.~P.~Abbott \textit{et al.} [LIGO Scientific and VIRGO],
%``GW170104: Observation of a 50-Solar-Mass Binary Black Hole Coalescence at Redshift 0.2,''
Phys. Rev. Lett. \textbf{118} (2017) no.22, 221101
%doi:10.1103/PhysRevLett.118.221101
[arXiv:1706.01812 [gr-qc]].

\bibitem{ligo4}
B.~P.~Abbott \textit{et al.} [LIGO Scientific and Virgo],
%``GW170814: A Three-Detector Observation of Gravitational Waves from a Binary Black Hole Coalescence,''
Phys. Rev. Lett. \textbf{119} (2017) no.14, 141101
%doi:10.1103/PhysRevLett.119.141101
[arXiv:1709.09660 [gr-qc]].

\bibitem{ligo5}
B.~P.~Abbott \textit{et al.} [LIGO Scientific and Virgo],
%``GW170608: Observation of a 19-solar-mass Binary Black Hole Coalescence,''
Astrophys. J. \textbf{851} (2017) no.2, L35
%doi:10.3847/2041-8213/aa9f0c
[arXiv:1711.05578 [astro-ph.HE]].

%old 
\bibitem{solutions} H.~Stephani, D.~Kramers, M.~A.~H.~MacCallum, C.~Hoenselaers, C.~Herlt, \textit{Exact solutions of Einstein's field equations}, Cambridge University Press (Cambridge, United Kingdom, 2003) .
 
\bibitem{SBH}
K.~Schwarzschild,
%``On the gravitational field of a mass point according to Einstein's theory,''
Sitzungsber. Preuss. Akad. Wiss. Berlin (Math. Phys. ) \textbf{1916} (1916), 189-196
[arXiv:physics/9905030 [physics]]. 
 
  
%\bibitem{RN} H. Reissner, Annalen Phys. 355 (1916) 106-120.

\bibitem{kerr}
R.~P.~Kerr,
%``Gravitational field of a spinning mass as an example of algebraically special metrics,''
Phys. Rev. Lett. \textbf{11} (1963), 237-238
%doi:10.1103/PhysRevLett.11.237

%%%%%%%%%%%%%%%%%%%%%%%%%%%%%%%%%%%%%%%%%%%%%%%%%%%%%%%

\bibitem{synge}
J.~L.~Synge,
%``The Escape of Photons from Gravitationally Intense Stars,''
Mon. Not. Roy. Astron. Soc. \textbf{131} (1966) no.3, 463-466
%doi:10.1093/mnras/131.3.463

\bibitem{luminet}
J.~P.~Luminet,
%``Image of a spherical black hole with thin accretion disk,''
Astron. Astrophys. \textbf{75} (1979), 228-235
 
\bibitem{Bambi:2008jg}
C.~Bambi and K.~Freese,
%``Apparent shape of super-spinning black holes,''
Phys. Rev. D \textbf{79} (2009), 043002
%doi:10.1103/PhysRevD.79.043002
[arXiv:0812.1328 [astro-ph]].  
  
\bibitem{Bambi:2010hf}
C.~Bambi and N.~Yoshida,
%``Shape and position of the shadow in the $\delta = 2$ Tomimatsu-Sato space-time,''
Class. Quant. Grav. \textbf{27} (2010), 205006
%doi:10.1088/0264-9381/27/20/205006
[arXiv:1004.3149 [gr-qc]].

\bibitem{study1}
A.~Abdujabbarov, F.~Atamurotov, Y.~Kucukakca, B.~Ahmedov and U.~Camci,
%``Shadow of Kerr-Taub-NUT black hole,''
Astrophys. Space Sci. \textbf{344} (2013), 429-435
%doi:10.1007/s10509-012-1337-6
[arXiv:1212.4949 [physics.gen-ph]].  

\bibitem{study2}
F.~Atamurotov, A.~Abdujabbarov and B.~Ahmedov,
%``Shadow of rotating Hořava-Lifshitz black hole,''
Astrophys. Space Sci. \textbf{348} (2013), 179-188
%doi:10.1007/s10509-013-1548-5

\bibitem{Moffat}
J.~W.~Moffat,
%``Modified Gravity Black Holes and their Observable Shadows,''
Eur. Phys. J. C \textbf{75} (2015) no.3, 130
%doi:10.1140/epjc/s10052-015-3352-6
[arXiv:1502.01677 [gr-qc]].
  
\bibitem{carlos1}
P.~V.~P.~Cunha, C.~A.~R.~Herdeiro, E.~Radu and H.~F.~Runarsson,
%``Shadows of Kerr black holes with scalar hair,''
Phys. Rev. Lett. \textbf{115} (2015) no.21, 211102
%doi:10.1103/PhysRevLett.115.211102
[arXiv:1509.00021 [gr-qc]].

\bibitem{quint2}
A.~Abdujabbarov, B.~Toshmatov, Z.~Stuchlík and B.~Ahmedov,
%``Shadow of the rotating black hole with quintessential energy in the presence of plasma,''
Int. J. Mod. Phys. D \textbf{26} (2016) no.06, 1750051
%doi:10.1142/S0218271817500511
[arXiv:1512.05206 [gr-qc]].

\bibitem{carlos2}
P.~V.~P.~Cunha, C.~A.~R.~Herdeiro, E.~Radu and H.~F.~Runarsson,
%``Shadows of Kerr black holes with and without scalar hair,''
Int. J. Mod. Phys. D \textbf{25} (2016) no.09, 1641021
%doi:10.1142/S0218271816410212
[arXiv:1605.08293 [gr-qc]].

\bibitem{study3}
Z.~Younsi, A.~Zhidenko, L.~Rezzolla, R.~Konoplya and Y.~Mizuno,
%``New method for shadow calculations: Application to parametrized axisymmetric black holes,''
Phys. Rev. D \textbf{94} (2016) no.8, 084025
%doi:10.1103/PhysRevD.94.084025
[arXiv:1607.05767 [gr-qc]].

\bibitem{study4}
P.~Cunha, V.P., C.~A.~R.~Herdeiro, B.~Kleihaus, J.~Kunz and E.~Radu,
%``Shadows of Einstein–dilaton–Gauss–Bonnet black holes,''
Phys. Lett. B \textbf{768} (2017), 373-379
%doi:10.1016/j.physletb.2017.03.020
[arXiv:1701.00079 [gr-qc]].

\bibitem{study5}
M.~Wang, S.~Chen and J.~Jing,
%``Shadow casted by a Konoplya-Zhidenko rotating non-Kerr black hole,''
JCAP \textbf{10} (2017), 051
%doi:10.1088/1475-7516/2017/10/051
[arXiv:1707.09451 [gr-qc]].

\bibitem{bobir2017}
B.~Toshmatov, Z.~Stuchlík and B.~Ahmedov,
%``Generic rotating regular black holes in general relativity coupled to nonlinear electrodynamics,''
Phys. Rev. D \textbf{95} (2017) no.8, 084037
%doi:10.1103/PhysRevD.95.084037
[arXiv:1704.07300 [gr-qc]].
  
\bibitem{study6}
H.~M.~Wang, Y.~M.~Xu and S.~W.~Wei,
%``Shadows of Kerr-like black holes in a modified gravity theory,''
JCAP \textbf{03} (2019), 046
%doi:10.1088/1475-7516/2019/03/046
[arXiv:1810.12767 [gr-qc]].

\bibitem{sudipta2019}
A.~K.~Mishra, S.~Chakraborty and S.~Sarkar,
%``Understanding photon sphere and black hole shadow in dynamically evolving spacetimes,''
Phys. Rev. D \textbf{99} (2019) no.10, 104080
%doi:10.1103/PhysRevD.99.104080
[arXiv:1903.06376 [gr-qc]].

\bibitem{shakih2019b}
R.~Shaikh,
%``Black hole shadow in a general rotating spacetime obtained through Newman-Janis algorithm,''
Phys. Rev. D \textbf{100} (2019) no.2, 024028
%doi:10.1103/PhysRevD.100.024028
[arXiv:1904.08322 [gr-qc]].

\bibitem{Konoplya:2019sns}
R.~A.~Konoplya,
%``Shadow of a black hole surrounded by dark matter,''
Phys. Lett. B \textbf{795} (2019), 1-6
%doi:10.1016/j.physletb.2019.05.043
[arXiv:1905.00064 [gr-qc]].

\bibitem{chicos}
E.~Contreras, J.~M.~Ramirez-Velasquez, Á.~Rincón, G.~Panotopoulos and P.~Bargueño,
%``Black hole shadow of a rotating polytropic black hole by the Newman–Janis algorithm without complexification,''
Eur. Phys. J. C \textbf{79} (2019) no.9, 802
%doi:10.1140/epjc/s10052-019-7309-z
[arXiv:1905.11443 [gr-qc]].

  
%%%%%%%%%%%%%%%%%%%%%%%%%%%%%%%%%%%%%%%%%%%%%%%%%%%%%%%%%%  

\bibitem{SN1}
A.~G.~Riess \textit{et al.} [Supernova Search Team],
%``Observational evidence from supernovae for an accelerating universe and a cosmological constant,''
Astron. J. \textbf{116} (1998), 1009-1038
%doi:10.1086/300499
[arXiv:astro-ph/9805201 [astro-ph]].

\bibitem{SN2}
S.~Perlmutter \textit{et al.} [Supernova Cosmology Project],
%``Measurements of $\Omega$ and $\Lambda$ from 42 high redshift supernovae,''
Astrophys. J. \textbf{517} (1999), 565-586
%doi:10.1086/307221
[arXiv:astro-ph/9812133 [astro-ph]].

\bibitem{adscft1}
J.~M.~Maldacena,
%``The Large N limit of superconformal field theories and supergravity,''
Int. J. Theor. Phys. \textbf{38} (1999), 1113-1133
%doi:10.1023/A:1026654312961
[arXiv:hep-th/9711200 [hep-th]].

\bibitem{adscft2}
I.~R.~Klebanov,
``TASI lectures: Introduction to the AdS / CFT correspondence,''
%doi:10.1142/9789812799630_0007
[arXiv:hep-th/0009139 [hep-th]].

\bibitem{BTZ1}
M.~Banados, C.~Teitelboim and J.~Zanelli,
%``The Black hole in three-dimensional space-time,''
Phys. Rev. Lett. \textbf{69} (1992), 1849-1851
%doi:10.1103/PhysRevLett.69.1849
[arXiv:hep-th/9204099 [hep-th]].

\bibitem{BTZ2}
M.~Banados, M.~Henneaux, C.~Teitelboim and J.~Zanelli,
%``Geometry of the (2+1) black hole,''
Phys. Rev. D \textbf{48} (1993), 1506-1525
%doi:10.1103/PhysRevD.48.1506
[arXiv:gr-qc/9302012 [gr-qc]].

\bibitem{lemos1}
J.~P.~S.~Lemos,
%``Cylindrical black hole in general relativity,''
Phys. Lett. B \textbf{353} (1995), 46-51
%doi:10.1016/0370-2693(95)00533-Q
[arXiv:gr-qc/9404041 [gr-qc]].

\bibitem{lemos2}
J.~P.~S.~Lemos and V.~T.~Zanchin,
%``Rotating charged black string and three-dimensional black holes,''
Phys. Rev. D \textbf{54} (1996), 3840-3853
%doi:10.1103/PhysRevD.54.3840
[arXiv:hep-th/9511188 [hep-th]].

\bibitem{lemos3}
V.~Cardoso and J.~P.~S.~Lemos,
%``Quasinormal modes of toroidal, cylindrical and planar black holes in anti-de Sitter space-times,''
Class. Quant. Grav. \textbf{18} (2001), 5257-5267
%doi:10.1088/0264-9381/18/23/319
[arXiv:gr-qc/0107098 [gr-qc]].

\bibitem{SD0}
Á.~Rincón, B.~Koch and I.~Reyes,
%``BTZ black hole assuming running couplings,''
J. Phys. Conf. Ser. \textbf{831} (2017) no.1, 012007
%doi:10.1088/1742-6596/831/1/012007
[arXiv:1701.04531 [hep-th]].

\bibitem{SD1}
B.~Koch, I.~A.~Reyes and Á.~Rincón,
%``A scale dependent black hole in three-dimensional space–time,''
Class. Quant. Grav. \textbf{33} (2016) no.22, 225010
%doi:10.1088/0264-9381/33/22/225010
[arXiv:1606.04123 [hep-th]].
  
\bibitem{SD2}
Á.~Rincón, E.~Contreras, P.~Bargueño, B.~Koch, G.~Panotopoulos and A.~Hernández-Arboleda,
%``Scale dependent three-dimensional charged black holes in linear and non-linear electrodynamics,''
Eur. Phys. J. C \textbf{77} (2017) no.7, 494
%doi:10.1140/epjc/s10052-017-5045-9
[arXiv:1704.04845 [hep-th]].
  
\bibitem{SD3}
Á.~Rincón and G.~Panotopoulos,
%``Quasinormal modes of scale dependent black holes in ( 1+2 )-dimensional Einstein-power-Maxwell theory,''
Phys. Rev. D \textbf{97} (2018) no.2, 024027
%doi:10.1103/PhysRevD.97.024027
[arXiv:1801.03248 [hep-th]].
  
\bibitem{SD4}
E.~Contreras, Á.~Rincón, B.~Koch and P.~Bargueño,
%``Scale-dependent polytropic black hole,''
Eur. Phys. J. C \textbf{78} (2018) no.3, 246
%doi:10.1140/epjc/s10052-018-5709-0
[arXiv:1803.03255 [gr-qc]].

\bibitem{SD5}
Á.~Rincón and B.~Koch,
%``Scale-dependent rotating BTZ black hole,''
Eur. Phys. J. C \textbf{78} (2018) no.12, 1022
%doi:10.1140/epjc/s10052-018-6488-3
[arXiv:1806.03024 [hep-th]].

\bibitem{SD6}
Á.~Rincón, E.~Contreras, P.~Bargueño, B.~Koch and G.~Panotopoulos,
%``Scale-dependent ( $2+1$ )-dimensional electrically charged black holes in Einstein-power-Maxwell theory,''
Eur. Phys. J. C \textbf{78} (2018) no.8, 641
%doi:10.1140/epjc/s10052-018-6106-4
[arXiv:1807.08047 [hep-th]].
 
 \bibitem{SD7}
F.~Canales, B.~Koch, C.~Laporte and A.~Rincon,
%``Cosmological constant problem: deflation during inflation,''
JCAP \textbf{01} (2020), 021
%doi:10.1088/1475-7516/2020/01/021
[arXiv:1812.10526 [gr-qc]].
  
\bibitem{SD8}
Á.~Rincón, E.~Contreras, P.~Bargueño and B.~Koch,
%``Scale-dependent planar Anti-de Sitter black hole,''
Eur. Phys. J. Plus \textbf{134} (2019) no.11, 557
%doi:10.1140/epjp/i2019-13081-5
[arXiv:1901.03650 [gr-qc]].

\bibitem{SD9}
E.~Contreras, Á.~Rincón, G.~Panotopoulos, P.~Bargueño and B.~Koch,
%``Black hole shadow of a rotating scale--dependent black hole,''
Phys. Rev. D \textbf{101} (2020) no.6, 064053
%doi:10.1103/PhysRevD.101.064053
[arXiv:1906.06990 [gr-qc]].

\bibitem{SD10}
G.~Panotopoulos, Á.~Rincón and I.~Lopes,
%``Interior solutions of relativistic stars in the scale-dependent scenario,''
Eur. Phys. J. C \textbf{80} (2020) no.4, 318
%doi:10.1140/epjc/s10052-020-7900-3
[arXiv:2004.02627 [gr-qc]].

%old
\bibitem{Rovelli:2007uwt} C.~Rovelli, 
{\it Philosophy of Physics},  
North Holland, 1287--1329 (2007).
  %``Quantum gravity,''
  
\bibitem{Stevenson:1981vj}
P.~M.~Stevenson,
%``Optimized Perturbation Theory,''
Phys. Rev. D \textbf{23} (1981), 2916
%doi:10.1103/PhysRevD.23.2916
  
\bibitem{Reuter:2003ca}
M.~Reuter and H.~Weyer,
%``Renormalization group improved gravitational actions: A Brans-Dicke approach,''
Phys. Rev. D \textbf{69} (2004), 104022
%doi:10.1103/PhysRevD.69.104022
[arXiv:hep-th/0311196 [hep-th]].
  
 \bibitem{Becker:2014qya}
D.~Becker and M.~Reuter,
%``En route to Background Independence: Broken split-symmetry, and how to restore it with bi-metric average actions,''
Annals Phys. \textbf{350} (2014), 225-301
%doi:10.1016/j.aop.2014.07.023
[arXiv:1404.4537 [hep-th]].

\bibitem{Dietz:2015owa}
J.~A.~Dietz and T.~R.~Morris,
%``Background independent exact renormalization group for conformally reduced gravity,''
JHEP \textbf{04} (2015), 118
%doi:10.1007/JHEP04(2015)118
[arXiv:1502.07396 [hep-th]].

\bibitem{Labus:2016lkh}
P.~Labus, T.~R.~Morris and Z.~H.~Slade,
%``Background independence in a background dependent renormalization group,''
Phys. Rev. D \textbf{94} (2016) no.2, 024007
%doi:10.1103/PhysRevD.94.024007
[arXiv:1603.04772 [hep-th]].

\bibitem{Morris:2016spn}
T.~R.~Morris,
%``Large curvature and background scale independence in single-metric approximations to asymptotic safety,''
JHEP \textbf{11} (2016), 160
%doi:10.1007/JHEP11(2016)160
[arXiv:1610.03081 [hep-th]].

\bibitem{Ohta:2017dsq}
N.~Ohta,
%``Background Scale Independence in Quantum Gravity,''
PTEP \textbf{2017} (2017) no.3, 033E02
%doi:10.1093/ptep/ptx020
[arXiv:1701.01506 [hep-th]].

\bibitem{BHreview}
P.~K.~Townsend,
``Black holes: Lecture notes,''
[arXiv:gr-qc/9707012 [gr-qc]].  

\bibitem{Biro:2017flp}
T.~S.~Bir{\'o}, V.~G.~Czinner, H.~Iguchi and P.~V{\'a}n,
%``Black hole horizons can hide positive heat capacity,''
Phys. Lett. B \textbf{782} (2018), 228-231
%doi:10.1016/j.physletb.2018.05.035
[arXiv:1712.09706 [gr-qc]].
  
\bibitem{btzRot}
C.~Martinez, C.~Teitelboim and J.~Zanelli,
%``Charged rotating black hole in three space-time dimensions,''
Phys. Rev. D \textbf{61} (2000), 104013
%doi:10.1103/PhysRevD.61.104013
[arXiv:hep-th/9912259 [hep-th]].
  
%old  
\bibitem{Riotto} A.~Riotto,
%``Inflation and the theory of cosmological perturbations,''
ICTP Lect. Notes Ser. \textbf{14} (2003), 317-413
[arXiv:hep-ph/0210162 [hep-ph]].   
  
\bibitem{Cai:1994np}
R.~G.~Cai, R.~K.~Su and P.~K.~N.~Yu,
%``Thermodynamics for black strings and p-branes,''
Phys. Lett. A \textbf{195} (1994), 307-311
%doi:10.1016/0375-9601(94)90034-5  
  
\bibitem{Gibbons:1976ue}
G.~W.~Gibbons and S.~W.~Hawking,
%``Action Integrals and Partition Functions in Quantum Gravity,''
Phys. Rev. D \textbf{15} (1977), 2752-2756
%doi:10.1103/PhysRevD.15.2752  
  
\bibitem{Kang:1996rj}
G.~Kang,
%``On black hole area in Brans-Dicke theory,''
Phys. Rev. D \textbf{54} (1996), 7483-7489
%doi:10.1103/PhysRevD.54.7483
[arXiv:gr-qc/9606020 [gr-qc]].


%%%%%%%%%%%%%%%%%%%%%%%%%%%%%%%%%%%%%%%%%%%

\bibitem{valeria}
V.~Ferrari and L.~Gualtieri,
%``Quasi-Normal Modes and Gravitational Wave Astronomy,''
Gen. Rel. Grav. \textbf{40} (2008), 945-970
%doi:10.1007/s10714-007-0585-1
[arXiv:0709.0657 [gr-qc]].

\bibitem{review1}
K.~D.~Kokkotas and B.~G.~Schmidt,
%``Quasinormal modes of stars and black holes,''
Living Rev. Rel. \textbf{2} (1999), 2
%doi:10.12942/lrr-1999-2
[arXiv:gr-qc/9909058 [gr-qc]].

\bibitem{review2}
E.~Berti, V.~Cardoso and A.~O.~Starinets,
%``Quasinormal modes of black holes and black branes,''
Class. Quant. Grav. \textbf{26} (2009), 163001
%doi:10.1088/0264-9381/26/16/163001
[arXiv:0905.2975 [gr-qc]].

\bibitem{review3}
R.~A.~Konoplya and A.~Zhidenko,
%``Quasinormal modes of black holes: From astrophysics to string theory,''
Rev. Mod. Phys. \textbf{83} (2011), 793-836
%doi:10.1103/RevModPhys.83.793
[arXiv:1102.4014 [gr-qc]].
  %
\end{thebibliography}
\end{document}